\documentclass[conference]{IEEEtran}
\usepackage{booktabs}
\usepackage{cite}
\usepackage{amsmath,amssymb,amsfonts}
\usepackage{algorithmic}
\usepackage{graphicx}
\usepackage{textcomp}
\usepackage{xcolor}
\usepackage{subfig}
\usepackage{booktabs}
\usepackage[export]{adjustbox}
\usepackage{tablefootnote}

\usepackage{svg}

\usepackage{xcolor}
\def\BibTeX{{\rm B\kern-.05em{\sc i\kern-.025em b}\kern-.08em
    T\kern-.1667em\lower.7ex\hbox{E}\kern-.125emX}}
\begin{document}

\vspace{-0.5cm}
\title{StutterNet: Stuttering Detection Using Time Delay Neural Network}

%

\author{\IEEEauthorblockN{Shakeel A.~Sheikh\textsuperscript{1}, Md Sahidullah\textsuperscript{1}, Fabrice Hirsch\textsuperscript{2}, Slim Ouni\textsuperscript{1}}
\IEEEauthorblockA{\textsuperscript{1}\textit{Universit\'{e} de Lorraine, CNRS, Inria, LORIA, F-54000, Nancy, France}\\
\textsuperscript{2}\textit{Université Paul-Valéry Montpellier}, CNRS, Praxiling, Montpellier, France
\\
\textsuperscript{1}\{shakeel-ahmad.sheikh, md.sahidullah, slim.ouni\}@loria.fr, \textsuperscript{2}fabrice.hirsch@univ-montp3.fr
}
\vspace{-1cm}
}

\maketitle

\begin{abstract}
This paper introduce~\emph{StutterNet}, a novel deep learning based stuttering detection capable of detecting and identifying various types of disfluencies. Most of the existing work in this domain uses automatic speech recognition (ASR) combined with language models for stuttering detection. Compared to the existing work, which depends on the ASR module, our method relies solely on the acoustic signal. We use a time-delay neural network (TDNN) suitable for capturing contextual aspects of the disfluent utterances. We evaluate our system on the \textit{UCLASS} stuttering dataset consisting of more than 100 speakers. Our method achieves promising results and outperforms the state-of-the-art residual neural network based method. The number of trainable parameters of the proposed method is also substantially less due to the parameter sharing scheme of TDNN.
\end{abstract}

\begin{IEEEkeywords}
stuttering, speech disfluency, speech disorder, time delay neural network.
\end{IEEEkeywords}
\vspace{-0.2cm}
\section{Introduction}
\label{introduction}

The \emph{speech disorders} problem refers to the difficulties in the production of speech sounds. The various speech disorders include \emph{cluttering}, \emph{lisping}, \emph{dysarthria}, \emph{stuttering}, etc. Of these speech disorders, stuttering -- also known as stammering -- is the most predominant one~\cite{guitar2013stuttering}. About 70 million people that comprise 1\% of the world population suffer from stuttering~\cite{b1}. People with the stuttering problem face several difficulties in social and professional interactions. This work is about the automatic detection of stuttering with several important applications. For example, it could facilitate the speech therapist's work, since they have to carry out a manual calculation to evaluate the severity of stuttering; to give a feedback to persons who stutter~(PWS) about their fluency. Nevertheless, fluent voice is an important requirement for several professions such as news anchoring, emergency announcement, etc. Furthermore, the automatic speech recognition (ASR) system used in voice assistants can be adapted efficiently for PWS.

Even though there are plenty of potential applications, stuttering detection has received less attention, especially from a signal processing and machine learning perspective. Stuttering is a neuro-developmental speech disorder, defined by an abnormally persistent and duration of stoppages in the normal forward flow of speech, which usually takes the form of \textit{core behaviors}: prolongations, blocks, and syllables, words or phrase repetitions \cite{guitar2013stuttering}. These impact the acoustic properties of speech which can help to discriminate from fluent voice. Studies show that different formant characteristics such as \emph{formant transitions}, \emph{formant fluctuations} are affected by stuttering~\cite{guitar2013stuttering}. The existing methods for stuttering detection employ spectral features such as \emph{mel-frequency cepstral coefficients} (MFCCs) and \emph{linear prediction cepstral coefficients} (LPCCs) or their variants that capture that formant-related information. Other spectral features such as pitch, zero-crossing rate, shimmer, and spectral spread are also used. Finally, those features are modeled with statistical modeling methods such as \emph{hidden Markov model} (HMM), \emph{support vector machine} (SVM), \emph{Gaussian mixture model} (GMM), etc~\cite{khara}.

An alternative strategy of stuttering detection is to apply ASR on the audio speech signal to get the spoken texts and then to use language models~\cite{alharbi, alharbi2, heeman}. Even though this method of detecting stuttering has achieved encouraging results and has been proven effective, the reliance on ASR makes it computationally expensive and prone to error. 

In this work, we use a deep neural network (DNN) for stuttering detection directly from the speech. In recent decades, the DNNs are widely used in different speech tasks such as speech recognition~\cite{speechrecog}, speaker recognition~\cite{latentspeech}, emotion detection~\cite{emotion},  voice disorder detection~\cite{disorder}. However, a little attention has been devoted to the field of stuttering detection.

We propose a \emph{time-delay neural network} (TDNN) architecture for stuttering detection. TDNN has been widely used for different speech classification problems such as speech and speaker recognition~\cite{tdnn,daniel}. We introduce this for stuttering detection task. The proposed method, referred to as \emph{StutterNet} is a multi-class classifier with output as stuttering types and fluent. Our experiments with the UCLASS dataset show promising recognition performance. We further optimize the \emph{StutterNet} architecture, and we achieved substantial improvement over the competitive DNN-based method.



\section{Related Work}
\vspace{-0.1cm}
\label{background}
The earlier studies in neural network based stuttering detection explored shallow architecture. Howell \emph{et al.}~\cite{howell1, howell2} employed two separate \emph{artificial neural networks} (ANNs) for the identification of repetition and prolongation disfluencies. This work used autocorrelation features, envelope parameters, and spectral information input to the neural network. The experiments were conducted with a dataset of 12 speakers. Ravikumar \emph{et al.}~\cite{ravi} attempted \emph{multilayer perceptron} (MLP) for the detection of repetition disfluencies. They used MFCC as input features from 12 disfluent speakers. I. Szczurowska \emph{et al.} employed Kohonen network and MLP for discriminating fluent and disfluent speech~\cite{szczurowska2014application}. The Kohonen network reduced the dimensionality of the Octave filter-based input feature. The features were used as an input to the MLP classifier. The experiments were conducted with eight speakers. B.~Villegas \emph{et al.}~\cite{villegas} proposed a respiratory-based stuttering classifier. They trained MLP on the respiratory air volume and pulse rate features for the detection of block stuttering. The network is trained on 68 Latin American Spanish speakers. The work in~\cite{manjula2019adaptive} used adaptive optimization based neural network for three class stuttering classification.

Due to recent advancements in deep learning, the improvement in speech technology surpasses the shallow neural network based approaches, and thus, resulted in a shift towards deep learning based framework and, disfluency identification is no exception.  The work in~\cite{oue2015automatic} used \emph{deep belief networks} with cepstral features for the detection of repetitions and stop gaps on TORGO dataset. T.~Kourkounakis \emph{et al}.~\cite{resnet_lstm} introduced a deep residual neural network and bi-directional long term short memory (ResNet+BiLSTM) based method to learn stutter-specific features from the audio. They addressed the stuttering detection problem as a multiple binary classification problem. They trained the same proposed architecture for each class of stuttering separately. The method was trained on 24 speakers from the UCLASS~\cite{uclass} stuttering datatset, and considered spectrograms as input feature. The learned features from residual blocks were fed to two bi-directional recurrent layers to capture the temporal context of the disfluent speech. 
\par 
Although this method has shown promising results in stuttering detection, it has several limitations. First, this method did not consider fluent speech, and the experiments are performed within stuttering classes. Second, the technique requires training of multiple models for each type of disfluency. Third, the model has a huge number of parameters ($\approx$24~million), thus makes it computationally expensive to train. Furthermore, the experiments are conducted with only a small subset of speakers.

In this paper, we address the above-mentioned problems with \emph{StutterNet} based on TDNN. This type of architecture is suitable for speech data as it captures temporal convolution as well as captures contextual information for a given context~\cite{tdnn}. We address stuttering type detection as a multi-class classification problem by training a single \emph{StutterNet} including data from all types of stuttering. Due to the parameter sharing in TDNN, we significantly reduce the number of parameters. In the next section, we provide the details of our proposed architecture. 
\vspace{-0.1cm}
\section{Proposed Architecture}
\label{pnetwork}
As discussed in Section \ref{introduction}, due to the very limited research in the field of stuttering detection, the idea is to design a single network that can be used to detect and identify various types of stuttering disfluencies. The proposed \emph{StutterNet} first computes the MFCC features from audio samples, which are then passed to the TDNN \cite{daniel} to learn and capture the temporal context of various types of disfluencies.

\subsection{Acoustic features}
In developing any speech domain application, the representative feature extraction is the most important that affects the model performance \cite{mahesha1, lschee2, resnet_lstm}. With the aid of signal processing techniques, several features of the stuttered speech signal can be extracted like raw waveform, spectrograms, mel-spectrograms, and or MFCCs. However, our aim is to compute and extract the features that compactly characterize the stuttering embedded in a speech segment and also which approximates the human auditory system’s response. For stuttering domain, MFCCs are the best suitable and are the most commonly used features in stuttered speech domain \cite{mfccs}, thus, we use MFCCs as the sole features to our \emph{StutterNet} network. These features are generated after every 12~ms on a 25~ms window for each 4~sec audio sample. This four-second window is used as stuttering lasts on average four seconds~\cite{guitar2013stuttering}.

\subsection{StutterNet architecture}
\begin{figure*}[]
\vspace{-.5cm}

\begin{minipage}[t]{0.7\linewidth}
    \centering
    \includegraphics[width=1\textwidth]{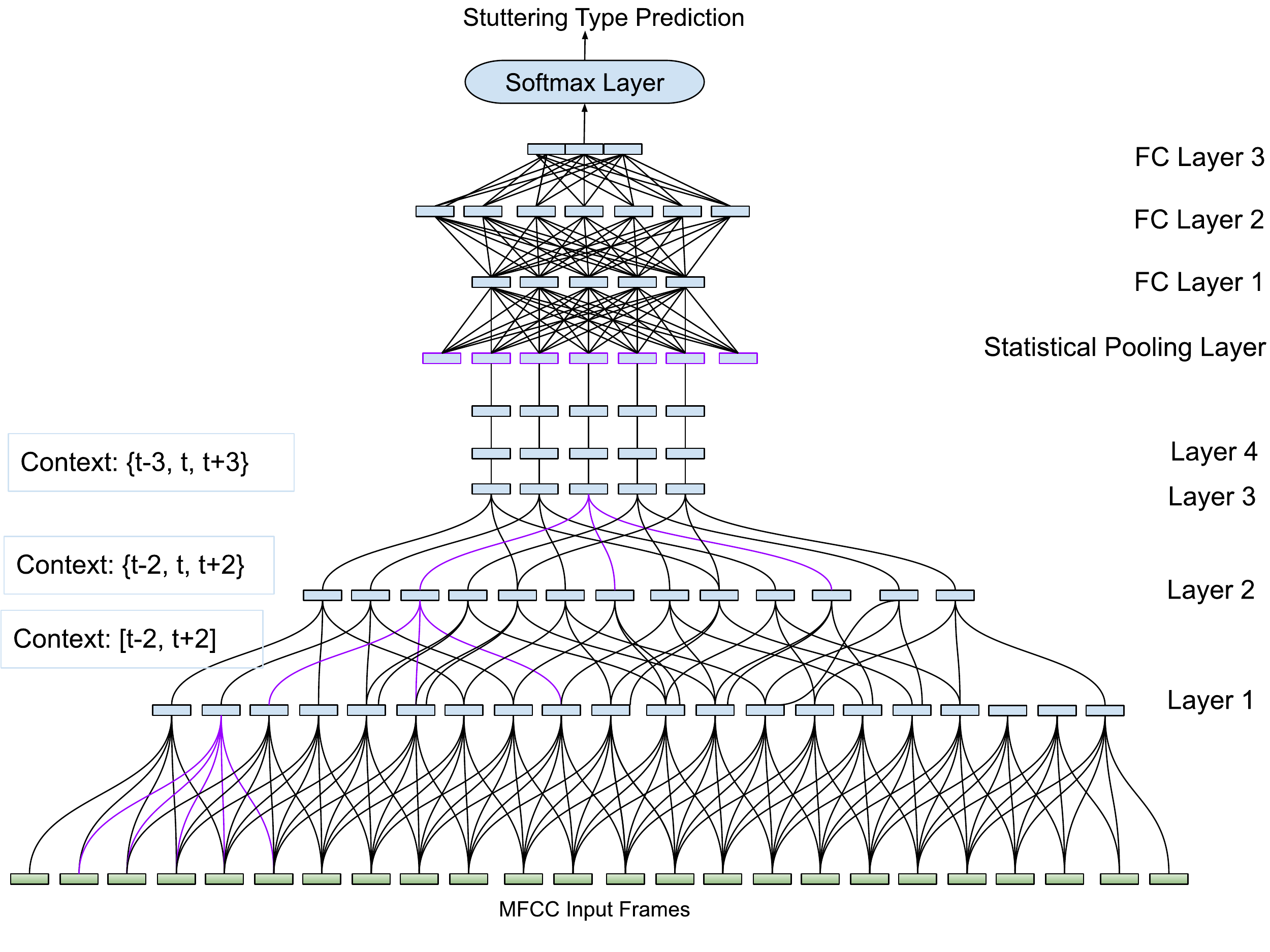}
    \vspace{-.5cm}

\end{minipage}
\begin{minipage}[t]{0.35\linewidth} 
    \centering
   \hspace*{-0.5cm} \includegraphics[scale=0.4]{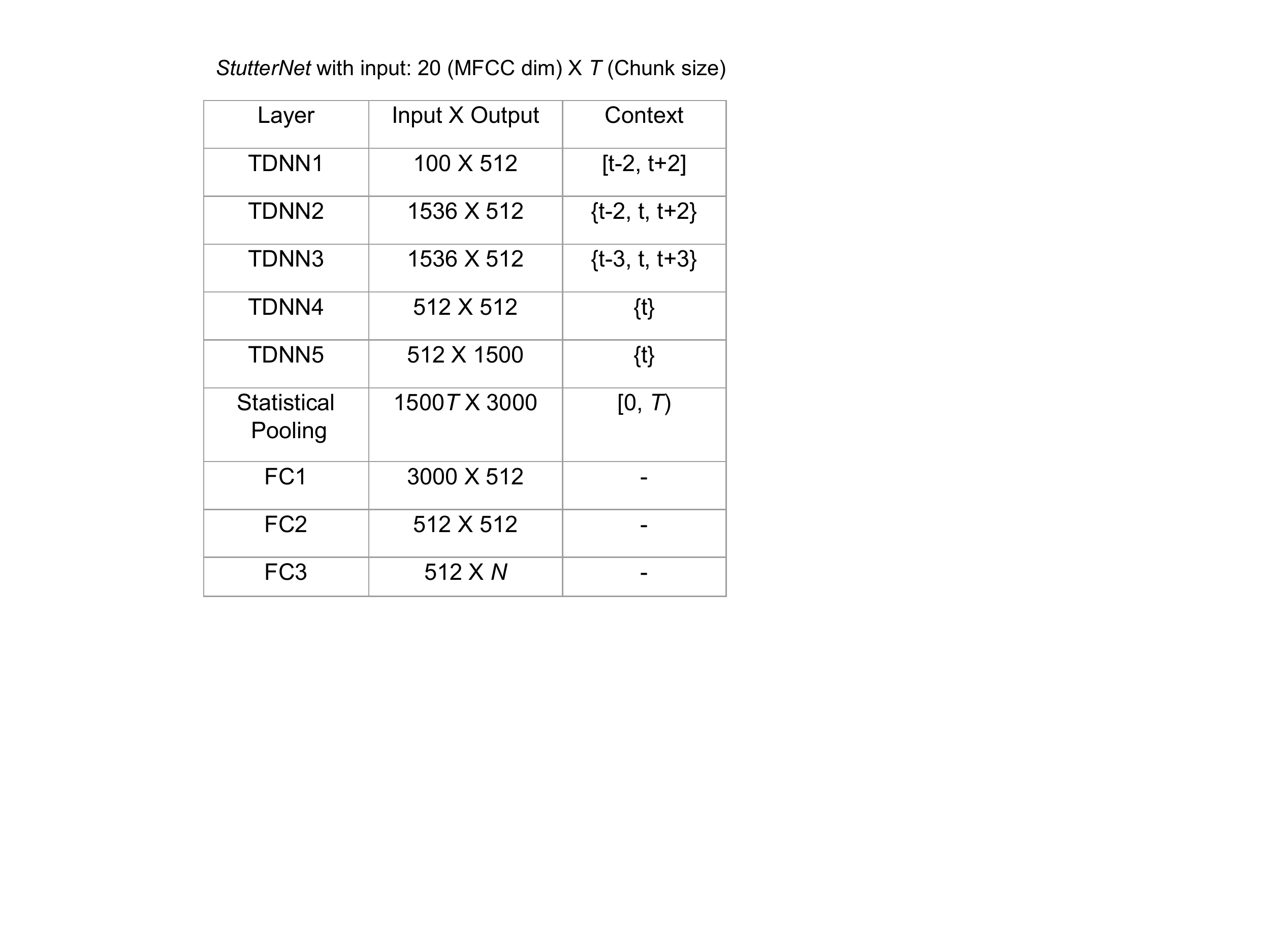}
    \vspace{-.5cm}

\end{minipage}   
\caption{\emph{(a)}:~StutterNet layers and context-wise computation,~\emph{(b)}:~\emph{StutterNet} Architecture (Baseline)~(Except statistical pooling layer, each layer is followed by a ReLU activation function and batch normalization)} 
\label{tdnn_architecture}
\vspace{-0.2cm}
\end{figure*}

Most of the existing work in literature has studied the stuttering detection as a binary classification problem: stuttering versus fluent detection \cite{khara} or one type versus other disfluency types \cite{resnet_lstm}. We tackle the stuttering detection as a multi-class problem of detecting and identifying the core behaviors, as opposed to the work done in \cite{resnet_lstm}, who addressed this problem as multiple binary classification, with the same network used for every disfluency. For this multi-class detection, we propose a TDNN \cite{tdnn}  based \emph{StutterNet} which effectively learns stutter-specific features. The TDNN method is well suited in capturing the temporal \cite{tdnn, daniel} and contextual aspects of various types of disfluencies.  The neural network takes 20 MFCCs as an input features to learn and capture the temporal context of stuttering. The \emph{StutterNet} contains five time delay layers with the first three focusing on the contextual frames of $[t-2,t+2], \{t-2,t,t+2\}, \{t-3,t,t+3\}$ with dilation of 1, 2 and 3 respectively. This is followed by statistical pooling, three fully connected (FC) layers and a softmax layer that reveals the prediction of multiclass stuttering disfluencies. Each layer is followed by a ReLU activation function and 1D batch normalization except statistical pooling layer. A dropout of 0.2 is applied to first two fully connected layers. The model architecture is shown in Fig.\ref{tdnn_architecture}.


\section{Experimental Evaluation}
\vspace{-0.1cm}
\label{experimentaldesign}
\subsection{Dataset}
We have used the UCLASS release 1 stuttering dataset that has been created by the Department of Psychology and Language Sciences, University College London~\cite{uclass}. This dataset consists of monologue samples from 139 participants aged between 5$-$45 years. Of these, 128 have been chosen in this case study with females 18 and males 110. Of these, 104 speakers were used for training, 12 for validation and 12 for testing.  The audio samples were annotated manually by listening to the recordings of speech segments. The annotations have been carried out into different categories of stuttering including: core behaviors, fluent, repetition$-$prolongations, blocks$-$prolongations, repetition$-$blocks, and blocks$-$repetition$-$prolongations. However, in this paper, we are focusing only on the core behaviors as the dataset contains only few of the remaining ones. Each monologue audio clip was sliced into 4-second and down sampled to 16 \textit{kHz} segments, resulting in a total of 4674 speech segment samples. Due to the lack of standard disfluent speech data, we have used only UCLASS dataset for our experimental studies. 
\par 
In order to evaluate the \emph{StutterNet} method on the UCLASS dataset, we adopted \textit{K}-fold cross validation technique , where \textit{K=10}. We conducted 10 experiments, each consisting of random sampling of 80\% for training, 10\% for validation and last 10\% for testing. The reported results  are the average between 10 experiments. All experiments were trained with an early stopping criteria of patience 7 on validation loss.

\subsection{Evaluation metrics}
In order to evaluate the model performance on this UCLASS dataset, we have used the metrics including:  precision, recall, F1-score and accuracy which are the standard and are widely used in the disfluent speech domain \cite{resnet_lstm}. In addition to these, we choose Matthew’s  correlation  coefficient (MCC), which is a balanced measure in the data imbalance problem~\cite{mcc}.
This measure lies between the range of $-1$ and $+1$. A value of $1$ represents the perfect prediction, $0$  is no better than the random guess and $-1$ shows total disagreement between observation and prediction. 
\vspace{-0.05cm}
\begin{equation}
    MCC = \frac{cs - \textbf{t}.\textbf{p}}{\sqrt{s^2 - \textbf{p}.\textbf{p}}\sqrt{s^2 - \textbf{t}.\textbf{t}}}
\end{equation}
where,
\begin{itemize}
    \item $s = \sum_{i}\sum_{j}C_{ij}$, is the total number of samples,
    \item $c = \sum_{k}C_{kk}$, is the total number of samples correctly predicted,
    \item $p_k = \sum_{i}C_{ki}$, is the number of times class k was predicted,
    \item $t_k = \sum_{i}C_{ik}$, is the number of times class k truly occurred,
\end{itemize}

\subsection{Implementation}
We develop \emph{StutterNet} with PyTorch library in Python~\cite{pytorch}. We use a learning rate of $10^{-4}$, amsgrad optimizer, and cross-entropy loss function. We use Librosa library~\cite{librosa} from Python for the feature extraction. We select models using an early stopping with a patience of seven epochs on validation loss. We compare the results obtained by \emph{StutterNet} to existing method~\cite{resnet_lstm} in the same experimental framework.

\begin{table*}[]
    \centering
    \renewcommand{\arraystretch}{1.1}
    \caption{Results in precision, recall and F1-score on UCLASS dataset (B: Block, F: Fluent, Rept: Repetition, Pr: Prolongation).}
    \vspace{-0.2cm}
    \begin{tabular}{*{14}{c}}
    \hline
    & \multicolumn{4}{c}{Precision} & \multicolumn{4}{c}{Recall} & \multicolumn{4}{c}{F1-Score} \\
    \hline
   Method &Rept&Pr&B&F&Rept&Pr&B&F&Rept&Pr&B&F\\
    \hline
    ResNet$+$BiLSTM~\cite{resnet_lstm} 
    &0.33&	0.42&	0.43&	\textbf{0.63}&	0.20&	\textbf{0.23}&	\textbf{0.53}&	0.55&	0.22&	\textbf{0.28}&	0.44&	0.52\\
   
    \hline
    \textbf{\emph{StutterNet}} (Baseline)
    &\textbf{0.36}&	\textbf{0.43}&	0.42&	0.59&	\textbf{0.28}&	0.17&	0.42&	0.67&	\textbf{0.30}&	0.23&	0.42&	0.62\\
    \textbf{\emph{StutterNet} (Optimized)}&
    0.35&	0.31&	\textbf{0.47}&	0.59&	0.24&	0.13&	0.47&	\textbf{0.70}&	0.27&	0.16&	\textbf{0.46}&	\textbf{0.63}\\
    \hline

    \end{tabular}

        \label{tab:results}
\end{table*}

\begin{table*}[]
    \centering
    \renewcommand{\arraystretch}{1.1}
    \caption{Results in accuracies and MCC on UCLASS dataset (B: Block, F: Fluent, Rept: Repetition, Pr: Prolongation).}
    \vspace{-0.2cm}
    \begin{tabular}{*{7}{c}}
    \hline
    \multicolumn{1}{c}{Method}&\multicolumn{4}{c}{Accuracy}&\multicolumn{1}{c}{Tot. Acc.}&\multicolumn{1}{c}{MCC.}\\
    \hline
    &Rept&Pr&B&F&&\\
    \hline
    Resnet+BiLSTM~\cite{resnet_lstm}&
    20.39&	\textbf{23.17}&	\textbf{53.33}&	55.00&	46.10&	0.20\\
    \hline
    \textit{ \textbf{StutterNet}  (Baseline)}&
    \textbf{27.88}&	17.13&	42.43&	66.63&	49.26&	0.21\\
    \textbf{\emph{StutterNet} (Optimized)}&
    23.98&	12.96&	47.14&	\textbf{69.69}&	\textbf{50.79}&	\textbf{0.23}\\
 \hline
    \end{tabular}
    
    \label{tab:Accuracy}
    \vspace{-0.2cm}
\end{table*}

\section{Results}
\vspace{-0.1cm}
The results of our baseline \emph{StutterNet} for different stuttering recognition are presented in Tables~\ref{tab:results} and \ref{tab:Accuracy}, where we compare our technique to ResNet+BiLSTM~\cite{resnet_lstm}. All the considered disfluencies and the fluent speech are  recognized with good scores. 
As can be seen from the Table~\ref{tab:results}, F1-Scores show clearly the good performances of baseline \emph{StutterNet} method in comparison to ResNet+BiLSTM. 
Table~\ref{tab:Accuracy} also shows that the baseline \emph{StutterNet} surpasses the state-of-the-art in most disfluency detection cases, but shows slightly lower performance in prolongation and block detection with an average accuracies of $17.13\%$, $42.43\%$ in comparison to $23.17\%$, $53.33\%$ average accuracies of ResNet+BiLSTM respectively. The \emph{StutterNet} outperforms ResNet+BiLSTM in correctly detecting fluent speech with a difference of 11.63 points ($66.63\%$ for the baseline, and $55.00\%$ for ResNet+BiLSTM).

\begin{figure}
    \centering
    \adjustbox{trim=1cm 0cm 1cm 0cm}{
    \includegraphics[scale=0.5]{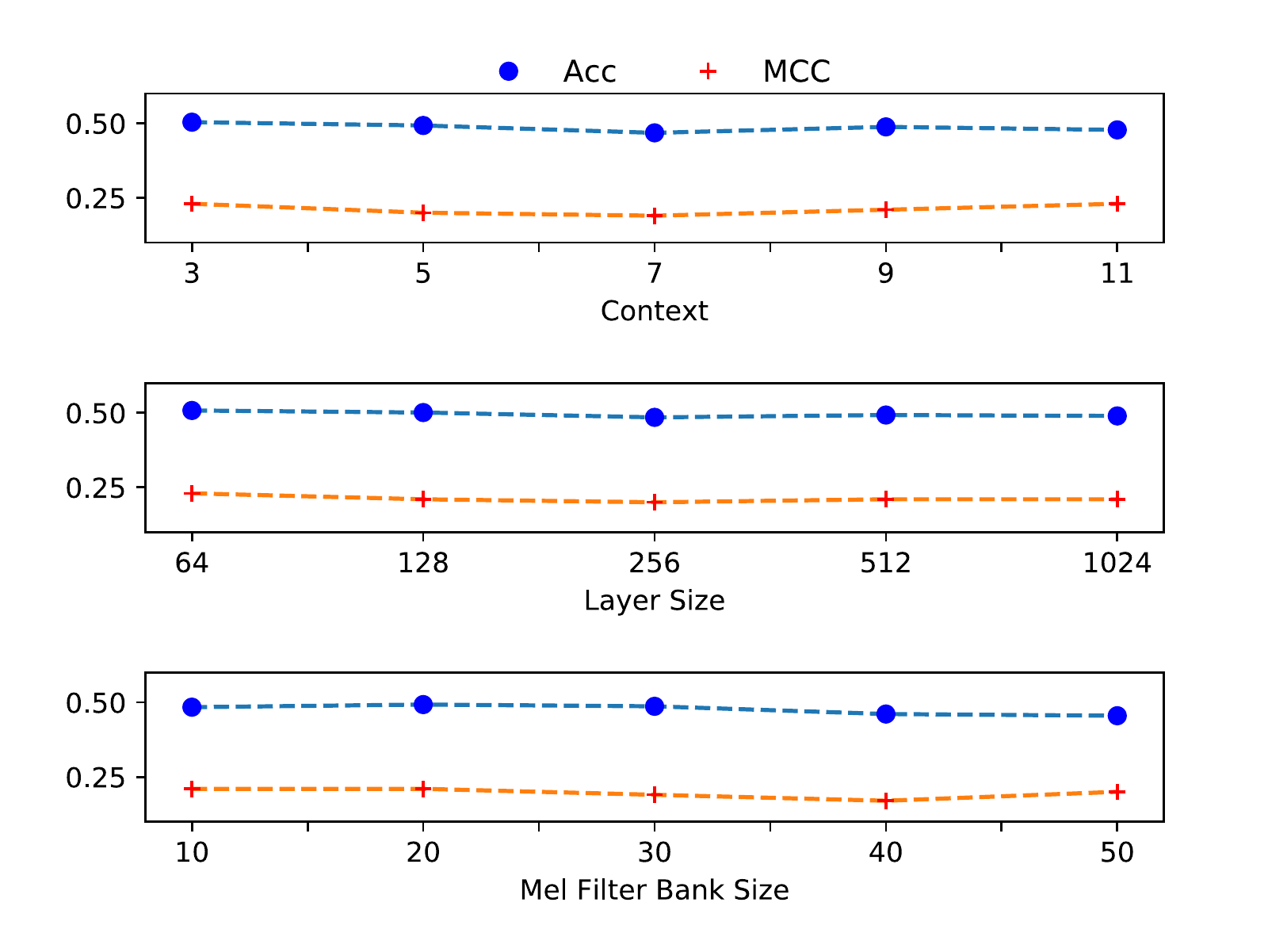}
    }
    \vspace{-0.2cm}
    \caption{\small MCC and accuracy of \emph{StutterNet} with varying context, layer size and mel filter bank size (Acc is normalized in [0,1]).}
    \label{fig:compare}
    \vspace{-0.7cm}
\end{figure}

We separately optimize the baseline \emph{StutterNet} by varying the filter bank size~(10,~30,~40,~50), context window~(3,~7,~9,~11) and layer size~(64,~128,~256,~1024). We found that context window optimization improves the detection performance of prolongation and repetition type of disfluencies. As the context increases, the performance of \emph{StutterNet} increases for the detection of prolongation and repetition stutterings, but decreases for the fluent segments, and it remains almost unchanged for the block stuttering. This makes sense because the repetition and prolongation disfluencies usually lasts longer and the longer context helps the \emph{StutterNet} in improving the performance. The block disfluencies doesn't last long: usually occurs at the beginning of speech segment and thus makes it context independent. We also found that layer size optimization slightly improves the performance (overall accuracy and MCC) of the stuttering detection in block and fluent types of disfluencies as shown in Fig.~\ref{fig:compare}. This might be due to the possible reason that the baseline \emph{StutterNet} is over-parameterized due to the limited size of the UCLASS dataset. We term the layer size optimized \emph{StutterNet} as optimized \emph{StutterNet} in Table~\ref{tab:results} and \ref{tab:Accuracy}. Compared to ResNet+BiLSTM, our optimized proposed method gains a margin of 4.69\% and 0.03 in overall average accuracy and MCC, respectively. For detecting the \emph{core behaviours} and the fluent part, the margins are also substantial (improvements of 7.49\%, 14.69\% in repetition and fluent speech segments, respectively).
Most previous work tends to avoid block disfluencies because of their similar nature to silence and prolongation (blocks are prolonged without audible airflow)~\cite{teesson2003lidcombe}. As shown in Table~\ref{tab:Accuracy}, the proposed \emph{StutterNet} can detect and classify the block stuttering with an average accuracy of 47.14\%. Moreover, our technique relies on the assumption that stuttering usually lasts for four-second window size~\cite{guitar2013stuttering}. Note that some of the stuttering~(in particular prolongation and repetition) can exceed more than four seconds in speech~\cite{guitar2013stuttering}, thus causing those prolongation stuttering likely to be misclassified. 

\par

\begin{figure*}[]
\vspace{-.5cm}

\begin{minipage}[t]{0.4\linewidth}
    \centering
    \includegraphics[width=1\textwidth]{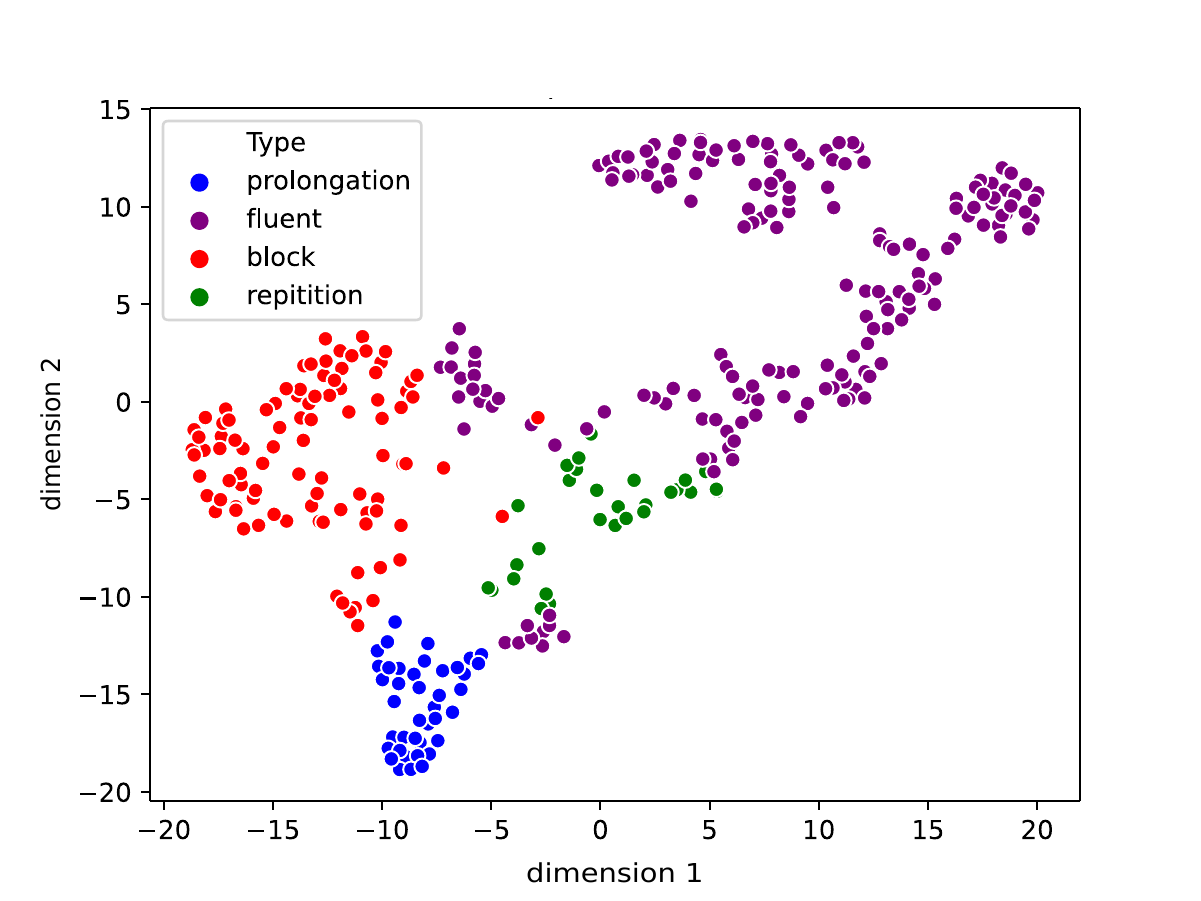}

\end{minipage}
\hspace{2cm}
\begin{minipage}[t]{0.4\linewidth} 
    \centering
    \includegraphics[width=1\textwidth]{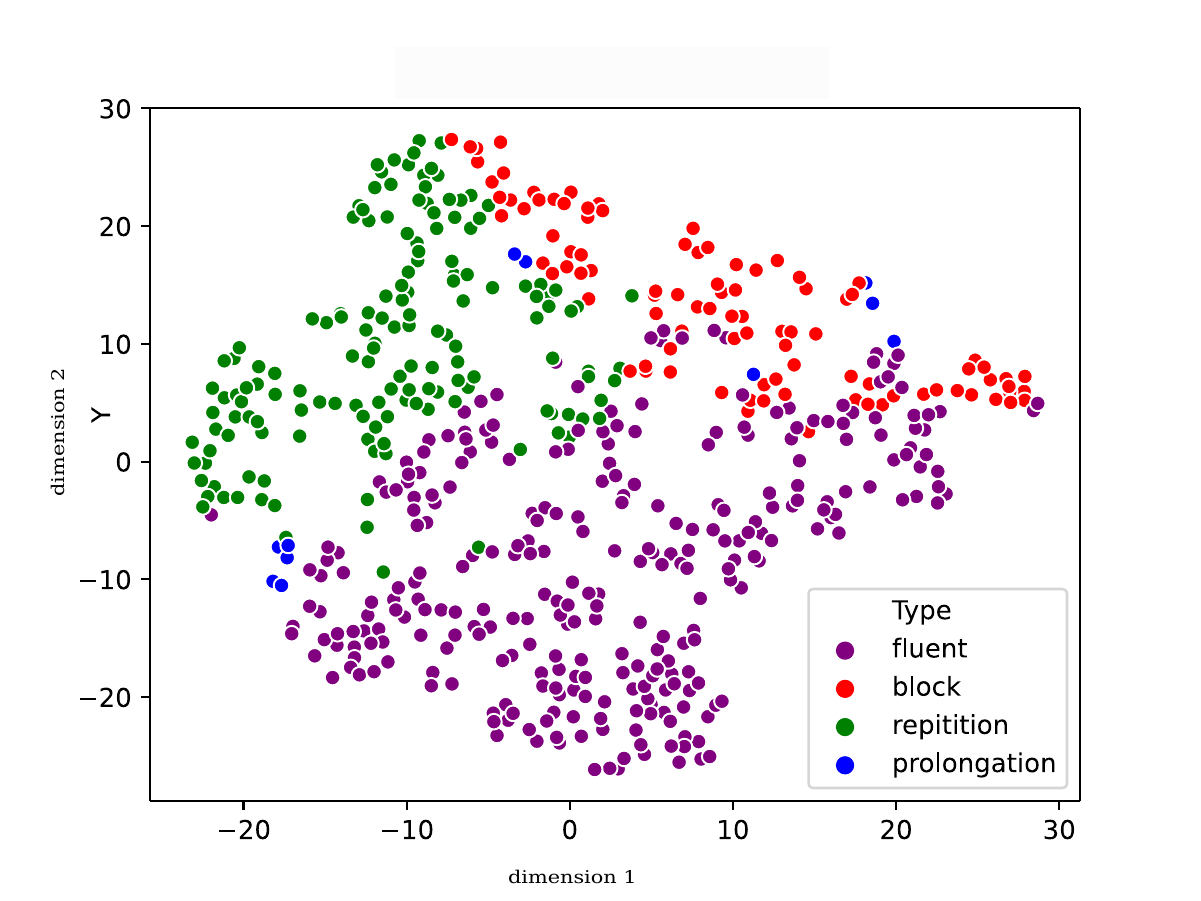}

\end{minipage}   
\vspace{-0.2cm}
\caption{t-SNE visualization of the output of last fully-connected layer for \emph{ResNet+BiLSTM} (left) and \emph{StutterNet} (right).}
\label{latentspace}
\vspace{-0.65cm}
\end{figure*}

Fig.~\ref{latentspace} shows a visualization of the latent feature embeddings learned by \emph{StutterNet} using t-SNE projection. Both ResNet+BiLSTM and \emph{StutterNet} present good discrimination of the different types of disfluencies. However, the latent feature embeddings learned by \emph{StutterNet} are more distinctive for fluent and less distinctive for prolongations and accurately capture the stutter-specific information than the state-of-the-art ResNet+BiLSTM method. Interestingly for ResNet+BiLSTM, the fluent and repetition category's embeddings are widely spread and more overlapped with the other classes. The prolongations and blocks are well clustered in BiLSTM+ResNet as compared to \emph{StutterNet}.

\vspace{-0.1cm}
\section{Conclusion}
\vspace{-0.1cm}
In this work, we present a \emph{StutterNet} to detect and classify several stuttering types. Our method uses a TDNN, which is trained on the MFCC input features. Only the \emph{core behaviors} and fluent part of the stuttered speech were considered in this study. The results show that the \emph{StutterNet} achieves considerable gain in overall average accuracy and MCC of 4.69\% and 0.03 respectively compared to the state-of-the-art method based on residual neural network and BiLSTM. We experimentally optimize layer size, context, and filterbank size in baseline \emph{StutterNet}. The performance moderately improves with layer size and context window optimization. Our method's main advantage is that it can detect all stuttering types with a single system with a smaller number of parameters, unlike the existing method. Our method also achieves considerable performance improvement in discriminating fluent vs. disfluent speech.

In this work, we have not evaluated the \emph{StutterNet} on the multiple disfluencies, where two or more disfluencies are present simultaneously in an utterance.  Besides, the \emph{UCLASS} dataset was collected in a controlled environment, whereas the real-time disfluency detection is a demanding problem. 
In future work, we will focus on multiple disfluencies by exploring the more advanced variants of TDNN for stuttering detection in a real-world scenario. We can also extend this work by exploring joint optimization of the different parameters, including context, filterbank size, and layer size of the proposed system.

\label{conc}



  \vspace{-0.2cm}
 \section*{Acknowledgment}
 \vspace{-0.2cm}
 This  work  was  made  with  the  support  of  the  French  National  Research Agency, in the framework of the project ANR BENEPHIDIRE (18-CE36-0008-03). Experiments  presented  in  this  paper  were  carried  out  using  the  Grid’5000 testbed, supported by a scientific interest group hosted by Inria and including CNRS,  RENATER  and  several  universities  as  well  as  other  organizations(see  https://www.grid5000.fr) and  using the EXPLOR  centre, hosted by the University of Lorraine.
 \vspace{-0.2cm}

\bibliographystyle{IEEEtran}
 \vspace{-0.1cm}

\bibliography{reference.bib}
 
\end{document}